\documentclass[aps,prb,twocolumn, superscriptaddress]{revtex4-1}
\usepackage[version=3]{mhchem} % Formula subscripts using \ce{}
\usepackage{amsmath} 
\usepackage{amssymb} 
\usepackage{textcomp}
\usepackage{graphicx}
\usepackage[utf8]{inputenc}
\usepackage{xcolor}
\begin{document}

\author{Craig M. Polley}
\email{craig.polley@gmail.com}
\affiliation{MAX IV Laboratory, Lund University, 221 00 Lund, Sweden}
\author{Ryszard Buczko}
\affiliation{Institute of Physics, Polish Academy of Sciences, 02-668 Warsaw, Poland}
\author{Alexander Forsman}
\affiliation{SCI Materials Physics, KTH Royal Institute of Technology, S-164 40 Kista, Sweden}
\author{Piotr Dziawa}
\author{Andrzej Szczerbakow}
\author{Rafa\l{} Rechci\'nski}
\author{Bogdan J. Kowalski}
\author{Tomasz Story}
\affiliation{Institute of Physics, Polish Academy of Sciences, 02-668 Warsaw, Poland}
\author{Ma\l{}gorzata Trzyna}
\affiliation{Center for Microelectronics and Nanotechnology, Rzeszow University, Rejtana 16A, Rzeszow 35-959, Poland}
\author{Marco Bianchi}
\author{Antonija Grubi{\v s}i{\'c} {\v C}abo}
\author{Philip Hofmann}
\affiliation{Department of Physics and Astronomy, Interdisciplinary Nanoscience Center (iNANO), Aarhus University, 8000 Aarhus C, Denmark}
\author{Oscar Tjernberg}
\affiliation{KTH Royal Institute of Technology, SCI Materials Physics, S-164 40 Kista, Sweden}
\author{Thiagarajan Balasubramanian}
\affiliation{MAX IV Laboratory, Lund University, 221 00 Lund, Sweden}

\title{Fragility of the Dirac Cone Splitting in Topological Crystalline Insulator Heterostructures}

%----------------- Abtract -------------------------

\begin{abstract}
\section*{Abstract}
The \lq{}double Dirac cone\rq{} 2D topological interface states found on the (001) faces of topological crystalline insulators such as Pb$_{1-x}$Sn$_{x}$Se feature degeneracies located away from time reversal invariant momenta, and are a manifestation of both mirror symmetry protection and valley interactions. Similar shifted degeneracies in 1D interface states have been highlighted as a potential basis for a topological transistor, but realizing such a device will require a detailed understanding of the intervalley physics involved. In addition, the operation of this or similar devices outside of ultra-high vacuum will require encapsulation, and the consequences of this for the topological interface state must be understood. Here we address both topics for the case of 2D surface states using angle-resolved photoemission spectroscopy. We examine bulk Pb$_{1-x}$Sn$_{x}$Se(001) crystals overgrown with PbSe, realizing trivial/topological heterostructures. We demonstrate that the valley interaction that splits the two Dirac cones at each $\bar{X}$ is extremely sensitive to atomic-scale details of the surface, exhibiting non-monotonic changes as PbSe deposition proceeds. This includes an apparent total collapse of the splitting for sub-monolayer coverage, eliminating the Lifshitz transition. For a large overlayer thickness we observe quantized PbSe states, possibly reflecting a symmetry confinement mechanism at the buried topological interface.
\\
\\
\end{abstract}
%----------------- Abtract -------------------------

\maketitle

The recent experimental realization of three dimensional topological insulators has triggered an expansive program of research, driven largely by the accessibility and rich physics of the 2D electronic states hosted on their surfaces.\cite{Ando2013,Qi2011,Bansil2016,Hasan2010,Zhang2012} These states are often said to be protected, referencing two distinct concepts. Firstly, the electrical connection of a band-inverted and normal material is not possible without the existence of interface states that thread the bulk bandgap. Secondly, in many band-inverted materials some kind of symmetry exists that forbids hybridization between these cross-gap interface states, thereby ensuring that they remain metallic. These basic characteristics of topological interface states are dictated solely by considerations of symmetry and bulk bandstructure, and in the sense that these are not easily affected by perturbations at the surface, can be considered robust. 

However it does not follow that the interface states are immune to surface perturbations. For example, adsorption of non-magnetic impurities on the topological insulator Bi$_2$Se$_3$ surface can dope the surface states.\cite{Benia2011,Bianchi2010,Valla2012} Similarly, the topological crystalline insulator (TCI) Pb$_{1-x}$Sn$_{x}$Se can be surface doped with Rb or Sn,\cite{Neupane2015,Pletikosic2014} while lattice distortions that disrupt mirror symmetry open a bandgap.\cite{Zeljkovic2015,Wojek2015} For both classes of materials, the surface termination is predicted to play an important role in the topological band dispersions.\cite{Zhu2014,Liu2013_prb,Wang2014,Safaei2013}

%------- FIGURE -------
\begin{figure}
	\includegraphics{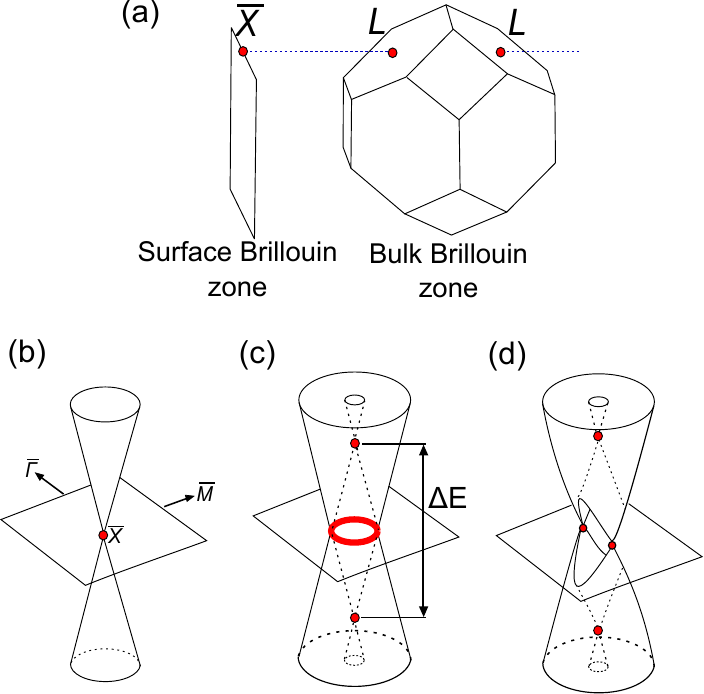}
	\caption{
		\textbf{Origin of the double Dirac cone on the Pb$_{1-x}$Sn$_{x}$Se (001) surface. (a) The bulk conduction-valence band inversion occurs at the L points in the bulk Brillouin zone. Two inequivalent L points are located at the same in-plane momentum as the surface $\bar{X}$ point. (b) The resulting topological surface state resembles a Dirac cone with two-fold valley degeneracy. (c) Intervalley coupling, the focus of this work, splits the degeneracy by $\Delta$E, yielding two Dirac cones that intersect in a circle. (d) Finally, hybridization occurs at all points in the intersection circle except for the two points lying in the mirror symmetry plane along $\bar{\Gamma}$-$\bar{X}$-$\bar{\Gamma}$.
		}
	}
	\label{fig:Origin}	
\end{figure}
%------- FIGURE -------
%------- FIGURE -------
\begin{figure*}
	\includegraphics{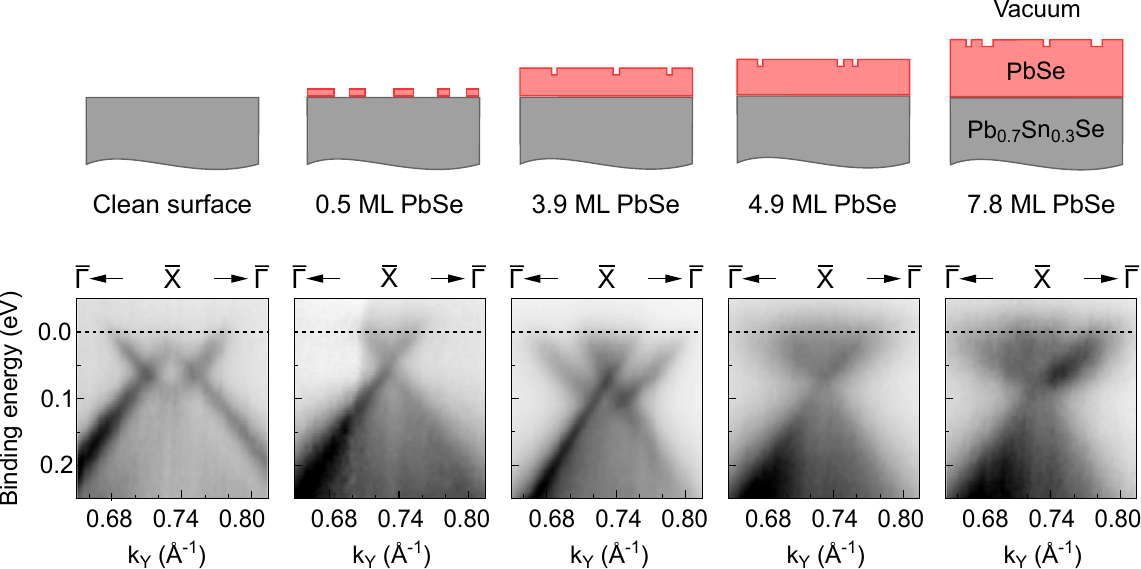}
	\caption{
		\textbf{Evolution of valley splitting in a TCI-normal heterostructure. Deposition of PbSe on the TCI Pb$_{1-x}$Sn$_{x}$Se buries the topological-normal interface responsible for the 2D electronic states, but for small thicknesses these states remain accessible to ARPES acquired at a resonant photon energy ($18$~eV). Sub-monolayer PbSe coverage is sufficient to collapse the valley splitting, while slightly more PbSe recovers it. Non-monotonic splitting can be tracked until blurring from inelastic photoelectron obscures the features of interest for PbSe thicknesses above 8 monolayers.
		}
	}
	\label{fig:PbSeDepth}	
\end{figure*}
%------- FIGURE -------
Details of the interface are also important in special cases where two bulk band inversions project to the same point in the surface Brillouin zone (Fig.\ref{fig:Origin}a), \textit{i.e.} they occur at the same in-plane momentum but different perpendicular momenta. Since it is uncommon for semiconductors to possess a direct band gap away from $\Gamma$, this condition is rarely satisfied. To date the only experimentally demonstrated examples are the (001) faces of (Pb,Sn)Se and (Pb,Sn)Te.\cite{Dziawa2012,Tanaka2012,Xu2012} A simple analysis would suggest that two degenerate Dirac-like surface states would derive from the two bulk band inversions (Fig.\ref{fig:Origin}b). In reality the degeneracy is lifted and the resulting bandstructure features crossings both at $\bar{X}$ and in a ring centered around $\bar{X}$ (Fig.\ref{fig:Origin}c). Hybridization eliminates all points of intersection except on the high symmetry line $\bar{\Gamma}$-$\bar{X}$-$\bar{\Gamma}$, where mirror symmetry forbids it. This results in the peculiar band dispersion shown in Fig. \ref{fig:Origin}d. 

The lifting of degeneracy depicted in Fig. \ref{fig:Origin}b (henceforth \lq{}valley splitting\rq{}) is the focus of this study. Valley splitting is an interface effect, and cannot be predicted purely from consideration of bulk properties. Symmetry analysis can produce a $k \cdot p$ Hamiltonian that accurately describes the surface bandstructure,\cite{Liu2013_prb,Wang2013} however this approach requires empirical tuning and leaves the physical interpretation of the Hamiltonian parameters unclear. Previous experimental studies have observed that the valley splitting depends on the bulk band-inversion magnitude,\cite{Wojek2014,Tanaka2013,Zeljkovic2015_2} but the mechanism remains to be clarified. 

Here we demonstrate an alternative approach to studying the valley splitting, in which we grow heterostructures consisting of thin layers of the normal semiconductor PbSe on the TCI Pb$_{0.7}$Sn$_{0.3}$Se. With angle resolved photoemission spectroscopy (ARPES) we are able to observe the evolution of the topological interface state as the normal/TCI interface is buried. To date there are very few spectroscopic studies of buried topological interface states.\cite{Berntsen2013,Yoshimi2014,Eich2016} We show that the magnitude of the valley splitting is extremely sensitive to atomic scale details of the interface, and moreover is non-monotonic with PbSe thickness (summarized in Fig.\ref{fig:PbSeDepth}). This evolution can be qualitatively explained with a simple effective mass model, accounting for both our results and those of previous studies. Tight binding calculations elucidate that the valley splitting decays with layer thickness, but in addition is strongly suppressed by PbSe steps at intermediate (non-integer) layer coverage.

%----------------------
\section*{Results and Discussion}
\subsection*{Physical Depth Profile}
%----------------------
%------- FIGURE -------
\begin{figure}
	\includegraphics{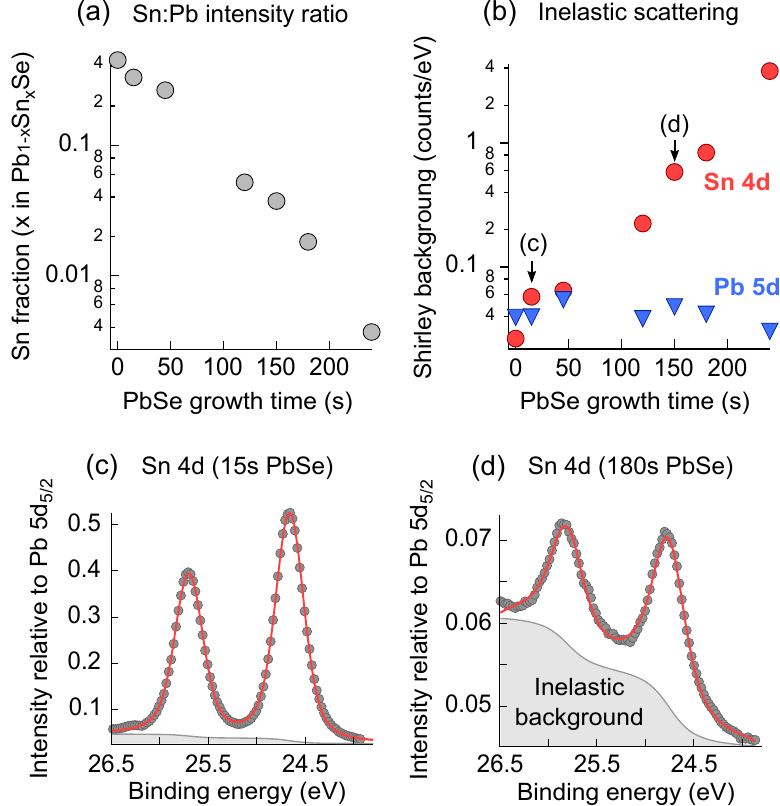}
	\caption{
		\textbf{Depth dependent core level spectroscopy. Monitoring the Pb5d and Sn4d core level spectra as a function of PbSe thickness provides important information about the growth. (a) If the Sn is being buried and not segregating, the Sn:Pb intensity ratio should and does exponentially decrease with overlayer thickness (owing to a Beer-Lambert absorption of outgoing photoelectrons). (b) By the same argument, the inelastic background of the Sn4d should and does increase while the Pb5d background remains constant. Analysis details can be found in the Supplementary Material. (c-d) Two Sn4d spectra illustrating both the intensity decrease and scattering increase)
		}
	}
	\label{fig:corelevels}	
\end{figure}
%------- FIGURE -------

Without leaving ultrahigh vacuum, synchrotron photoemission measurements were performed on the heterostructure samples. Monitoring the intensity ratio of the Sn 4d and Pb 5d core levels (Fig. \ref{fig:corelevels}a) offers an \textit{in-situ} estimate of the growth rate, since lead atoms remain at the surface while tin atoms are buried.  As described in the Supplementary Material, this method returns a growth rate of $\approx$0.08~\AA s$^{-1}$. This can be corroborated by \textit{ex-situ} measurements: secondary ion mass spectrometry (SIMS) on the same sample suggests a rate of $\approx$0.11~\AA s$^{-1}$, while cross sectional electron microscopy of a thick calibration film indicates a growth rate of (0.10 $\pm$ 0.01)~\AA s$^{-1}$ (see Supplementary Material). Since electron microscopy is the best calibrated approach, in the remainder of the manuscript we will assume this growth rate when mapping from growth time to film thickness in monolayers (1~ML = 0.306~nm). None of the techniques indicated thickness variation across the sample laterally.

Two important assumptions in these experiments are that the PbSe / Pb$_{0.7}$Sn$_{0.3}$Se interface is sharp and that the band ordering of the PbSe layer is normal. Here the two major concerns are segregation of tin atoms and strain of the PbSe layer, since both could result in a band inverted PbSe overlayer. Owing to the low growth temperatures, segregation seems unlikely and indeed the exponentially decreasing core level intensity ratio in Fig \ref{fig:corelevels}a suggests that tin atoms remain at the interface. An independent confirmation of this can be obtained from examining the strength of the Shirley background of the core level spectra (Fig. \ref{fig:corelevels}b-d). As a function of PbSe thickness, the Pb 5d background is unchanged while the Sn 4d background increases exponentially. This is consistent with the Sn atoms remaining buried at the interface, resulting in increased scattering of the photoelectrons as they travel to the surface. From our measurements it is not possible to evaluate strain in the PbSe layer. The lattice constant of Pb$_{0.7}$Sn$_{0.3}$Se (a=6.08~\AA) is slightly smaller than that of PbSe (a=6.12~\AA), but the thermal expansion coefficients are essentially identical. Hence if no relaxation occurs during growth, biaxial compressive strain of at most 0.7\% could be expected. Previous calculations \cite{Barone2013, Hsieh2012} predict that PbSe requires compressive strain of approximately 2\% in order to close the bandgap. We might therefore expect the PbSe bandgap to be reduced, but not closed and certainly not inverted. Due to this low level of strain and film thicknesses below 10nm, we also do not expect the periodic dislocation network seen previously in STM experiments on IV-VI heterostructures with larger lattice mismatches ($>$3\%).\cite{Springholz2001,Zeljkovic2015_2}

\subsection*{Bandstructure Measurements}
\begin{figure*}
	\includegraphics{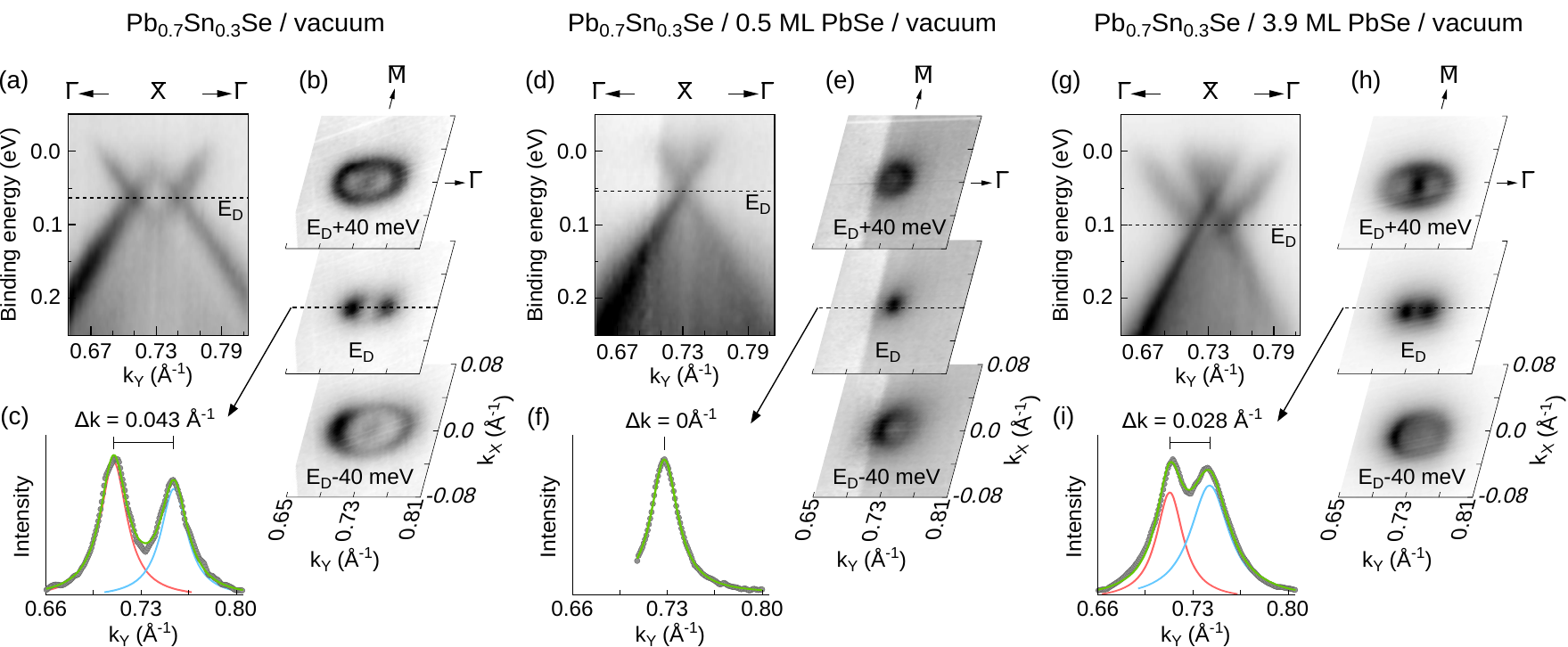}
	\caption{
		\textbf{Momentum resolved spectra of the evolving TCI-normal interface state. The dispersions for (a-c) the clean vacuum / Pb$_{0.7}$Sn$_{0.3}$Se, (d-f) the same after depositing 0.5~ML of PbSe and (g-i) after depositing a further 3.4~ML of PbSe. Constant energy cuts (b,e,h) clearly demonstrate the varying momentum splitting of the child Dirac points, a direct consequence of the varying energy splitting of the parent Dirac cones. Momentum distribution curves through the child Dirac points (c,f,i) quantify the momentum splitting for the three thickness steps. The stepwise change in intensity in (d,e) is an intensity normalization artifact and is the reason for the truncation in (f).
		}
	}
	\label{fig:Gap}	
\end{figure*}

In Fig.\ref{fig:Gap} we show ARPES datasets measured at three different PbSe overlayer thicknesses. The clean surface (Fig. \ref{fig:Gap}a-c) shows the well known double Dirac cone structure, where the energy splitting of two \lq{}parent\rq{} Dirac cones results in the emergence of two momentum split \lq{}child\rq{} Dirac points away from $\bar{X}$. After adding only 0.5~ML of PbSe, the valley splitting of the parent Dirac cones collapses (Fig. \ref{fig:Gap}d-f), taking with it the characteristic Lifshitz transition and associated van-Hove singularities. When an additional 3.4~ML are deposited, the splitting returns (Fig. \ref{fig:Gap}g-i). Note that the constant energy surfaces in Fig. \ref{fig:Gap}h clearly indicate two Dirac points, excluding the possibility that this is a single Dirac cone plus a trivial parabolic conduction band state. Careful inspection reveals that the k$_\parallel$ dispersion has also changed; the simple $k \cdot p$ model of Liu \cite{Liu2013_prb} includes such behaviour provided we include higher order terms (discussed in the Supplementary Material). Beyond this thickness the splitting appears to collapse once more, although it becomes difficult to be certain owing to the unavoidable increase in blurring of the spectra from inelastic photoelectron scattering. Several additional growth experiments, summarized in Supplementary Material, show a similar evolution. The striking changes in the surface state dispersion with increasing PbSe thickness constitute the key result of this study, and the discussion that follows is directed at understanding the origin of this behaviour. 

The choice of 18~eV as photon energy for the ARPES measurements is significant, partially for high momentum resolution but primarily due to a strong resonant enhancement of the interface state at this energy (also seen on the (111) oriented surface \cite{Polley2014}). The inelastic mean free path for photoelectrons is very short at this energy ($<$2~nm) and we are attempting to measure electronic states localized to a buried interface. However the combination of the exponential decay of the interface state wavefunction towards the surface, the very strong intensity enhancement at h$\nu$=18~eV and the high photon flux available from a synchrotron source means that the topological interface state can indeed be observed for PbSe layer thicknesses up to at least 8~ML (2.5~nm). A similar approach has enabled the measurement of buried dopant layers in silicon.\cite{Miwa2013,Miwa2014}

\subsection*{Thickness Dependent Energy Splitting}
%----------------------
%------- FIGURE -------
%------- FIGURE -------
To understand the Dirac cone splitting, useful comparisons may be drawn with the valley splitting of quantum well states hosted by semiconductors such as Si,\cite{AndoFowlerStern1982,Zwanenberg2013} Ge \cite{Carter2013} and AlSb.\cite{Ting1988} The effective mass approximation (EMA) is a useful framework here. Under the EMA, eigenstates of a confinement potential at a given momentum are constructed from a basis of all bulk Bloch states with the same in-plane momentum but various perpendicular momenta. The resulting wavefunction can be approximated by a simple envelope function, abstracting away the lattice potential and reducing the problem to that of a simpler quantum mechanical \lq{}particle in a box\rq{} solution to the bare confinement potential. Several authors have demonstrated that the EMA remains valid with a two-valley bulk basis,\cite{Sham1979,Ting1988,Valavanis2007,Saraiva2011,Nestoklon2006} in which case two degenerate eigenstates are predicted. However experimentally it has long been known that the degeneracy is usually lifted.\cite{AndoFowlerStern1982} The reason is that the two eigenstates will be built from the bulk basis such that they have the same envelope function but differing phases of the underlying Bloch-like oscillation. In free space this has no consequence, but in the presence of a confinement or interface potential the energy eigenvalues slightly differ. In this way the situation is closely analogous to the opening of bandgaps at zone boundaries in the nearly-free-electron model. 

Accurately calculating the splitting magnitude is challenging, but certain qualitative aspects are universally agreed on. The magnitude of the splitting is typically small, but increases rapidly with the strength of the confinement potential. It also oscillates as the width of the confinement potential changes on an atomic scale,\cite{Ting1988,Nestoklon2006,Valavanis2007,Freisen2007,Boykin2008,Virgilio2009} reflecting the phase difference between the two eigenstates. Finally, and for similar reasons, the presence of disorder or steps can strongly suppress valley splitting.\cite{Goswami2007,Zwanenberg2013} 

The physics described here is independent of the material system or the origin of the confinement potential. It applies to the two-dimensional states formed by electrostatic band bending in field effect transistor inversion layers, semiconductor heterostructures and delta-doping profiles, but also to the mass inversions at the boundaries of topological insulators. However certain properties are unique to band-inverted heterostructures. The wavefunction is peaked at the well walls rather than inside the well, with an envelope function that decays exponentially as:\cite{Zhang2012}

\begin{equation}\label{eqn:F0}
\phi(z) = Ae^{-\frac{|E_g|}{2v}z}
\end{equation}

where $A$ is a scalar normalization factor,  $E_g$ the bulk bandgap and $v$ the bulk band velocity in the $z$ direction. Hence to influence the real-space confinement of the wavefunction, it is the magnitude and abruptness of the \textit{band inversion} that must be tuned. In contrast to regular quantum wells, electric fields are ineffective to tune the envelope function, since this moves the energy positions of valence and conduction bands equally. This can already be seen in the results of previous studies: electrostatic surface doping of (Pb,Sn)Se causes a rigid energy shift but has no effect on the valley splitting,\cite{Pletikosic2014,Neupane2015} while increasing the band-inversion strength in the bulk by altering composition, temperature or strain increases the valley splitting .\cite{Wojek2014,Tanaka2013,Zeljkovic2015_2} The latter case can be interpreted as pushing the surface state wavefunction towards the strong potential gradient at the TCI-vacuum interface, thus making the effect of the phase difference more pronounced.

\begin{figure}
	\includegraphics{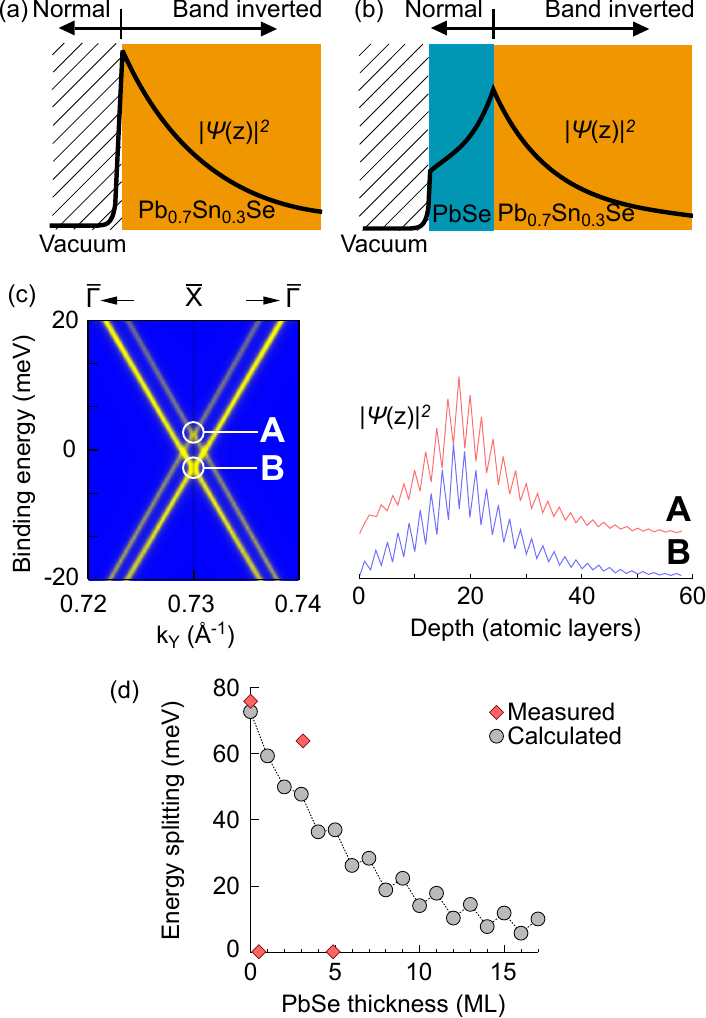}
	\caption{
		\textbf{Origin of valley splitting. Sketches of the interface state wavefunction envelope in the case of (a) a vacuum / Pb$_{0.7}$Sn$_{0.3}$Se structure and (b) a vacuum / PbSe / Pb$_{0.7}$Sn$_{0.3}$Se heterostructure. Since PbSe has normal band ordering with a finite bandgap, it serves to slow the wavefunction decay on the normal side. Tight binding calculations (c) confirm the overall shape of the wavefunction envelope, as well as the differing phases of the fast oscillations that are responsible for the valley splitting. Extended calculations as a function of the PbSe overlayer thickness (d, grey circles) confirm an overall decay of the valley splitting modulated by rapid oscillations. Experimental measurements (d, red diamonds) however show a much stronger dependence on thickness.
		}
	}
	\label{fig:Oscillation}	
\end{figure}

In Fig.\ref{fig:Oscillation}a,b we sketch the situation for Pb$_{0.7}$Sn$_{0.3}$Se with a PbSe overlayer. Considering Equation \ref{eqn:F0} in the limit of a clean surface, the envelope function decays slowly into the Pb$_{0.7}$Sn$_{0.3}$Se but abruptly into the vacuum, since vacuum can be considered equivalent to a trivial insulator with infinite bandgap. In contrast, PbSe is a finite gap trivial insulator and establishes a slower, piecewise decay of the envelope on the trivial side (Fig. \ref{fig:Oscillation}b). As the thickness of the PbSe layer increases, the interface state wavefunction grows wider and the envelope amplitude at both the PbSe/(Pb,Sn)Se and vacuum/PbSe interfaces decreases. In line with the preceding discussion, we should expect this to result in an overall decrease of the valley splitting, accompanied by two-monolayer periodic oscillations.  Tight binding calculations of PbSe / Pb$_{0.7}$Sn$_{0.3}$Se heterostructures make a similar prediction. Fig.\ref{fig:Oscillation}c plots the surface spectral function for a 16~ML thick PbSe overlayer on Pb$_{0.7}$Sn$_{0.3}$Se(001), and demonstrates a small energy splitting. The depth-resolved probability densities at k$_\parallel=\bar{X}$ are in agreement with the preceding discussion, with envelope amplitudes peaked at the PbSe / Pb$_{0.7}$Sn$_{0.3}$Se interface and decaying exponentially on either side. The antiphase of the underlying fast oscillations is also apparent. Performing the same calculations as a function of PbSe thickness yields the expected depth-dependent reduction in splitting shown in Fig.\ref{fig:Oscillation}d (grey circles).  

%------- FIGURE -------

%------- FIGURE -------

However when these calculations are compared with our experimental results (Fig.\ref{fig:Oscillation}d, red diamonds), it is clear that we observe much larger changes in energy splitting. In particular, without even a full monolayer coverage of PbSe the experimental splitting has apparently completely collapsed. This can be accounted for by considering the additional influence of terraces for non-integer layer coverage. 

\subsection*{The Role of Atomic Terraces}
\begin{figure}
	\includegraphics{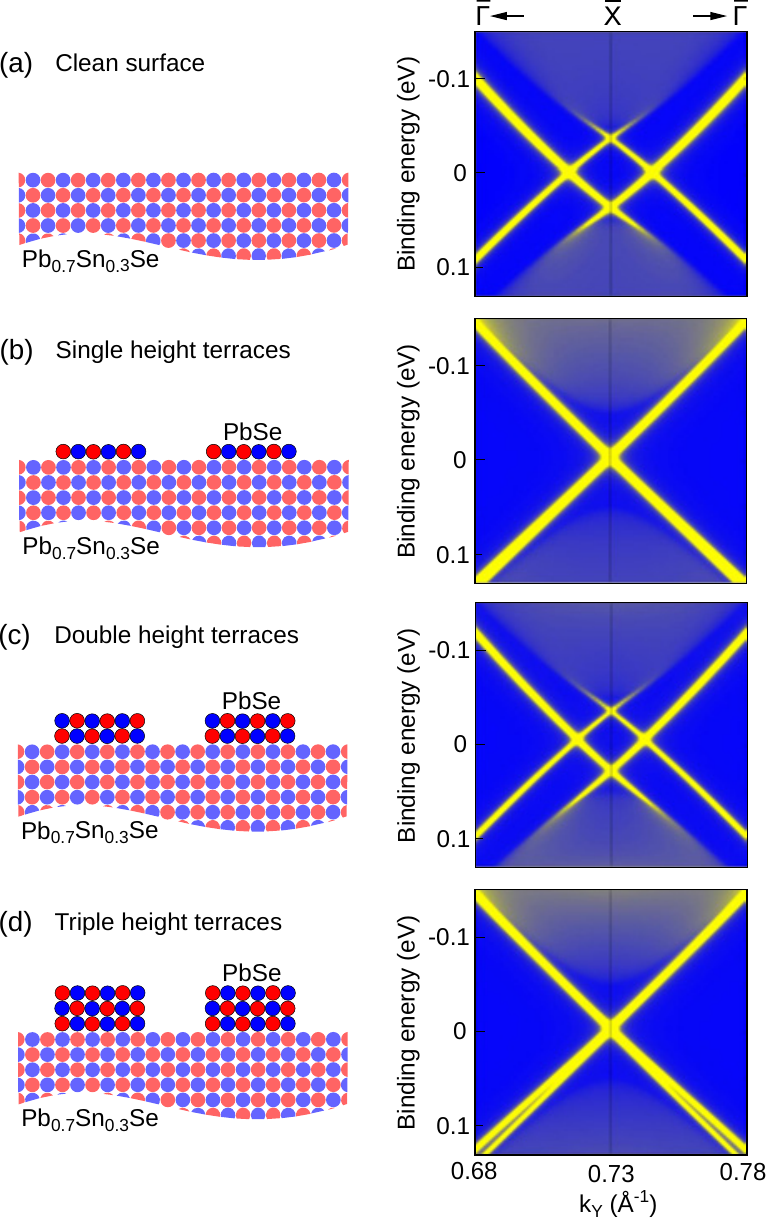}
	\caption{
		\textbf{Influence of atomic terraces. Surface spectral functions in the presence of a regular array of atomic terraces, corresponding to incomplete layer coverage. The calculations use terraces 31 atoms wide with a 50 atom period, and aligned along the [110] direction. Terraces with odd atomic heights (b,d) strongly suppress the valley splitting due to the resulting phase difference between the two eigenstates. Even height terraces (c) recover nearly the same splitting as the flat surface (a).
		}
	}
	\label{fig:Terraces}	
\end{figure}

Additional tight binding calculations (Fig. \ref{fig:Terraces}) illustrate the pronounced effect of surface terraces for the case of partial PbSe coverage. This is achieved by constructing an in-plane structure of 31 atom wide PbSe terraces, aligned to the [110] direction and separated by 19 atoms. This coverage corresponds to 0.62~ML, approximately matching the collapsed Dirac cones in Fig. \ref{fig:Gap}d-f. For odd height terraces there is an almost complete collapse of the splitting, while for even height terraces the splitting is largely unaffected. This is a direct consequence of the antiphase of the two eigenstates, as we discuss in detail in the Supplementary Material. In simple terms if the eigenstates are in antiphase, a shift of the surface position equivalent to a $\pi$ phase shift should exchange bonding and antibonding states. For IV-VI films this occurs every atomic layer. As a result, if the surface is half covered by monoatomic terraces then any fixed phase configuration will integrate to zero splitting.

%------- FIGURE -------
%------- FIGURE -------

%------- FIGURE -------
\begin{figure*}
	\includegraphics{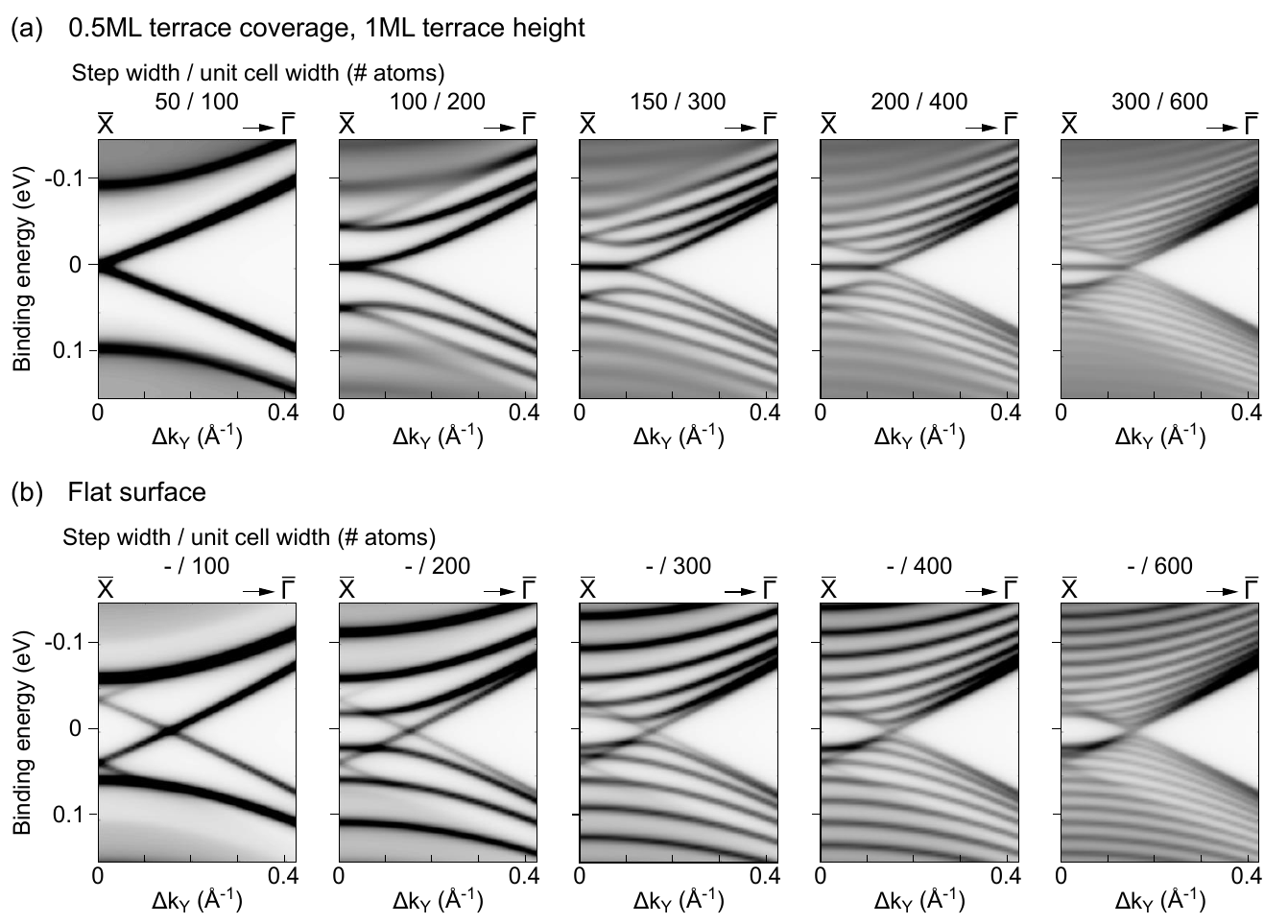}
	\caption{
		\textbf{Size dependence of splitting cancellation. Spectral density functions for a Pb$_{0.7}$Sn$_{0.3}$Se surface with 0.5~ML coverage of Pb$_{0.7}$Sn$_{0.3}$Se terraces aligned along the [110] direction. For sufficiently large terrace sizes the splitting cancellation no longer occurs (a). Repeating the calculations with a flat surface (b) illustrates the contribution of band-folding due to the increasingly large unit cell. Note that for compactness only $\bar{X}$-$\bar{\Gamma}$ sections are plotted here as opposed to $\bar{\Gamma}$-$\bar{X}$-$\bar{\Gamma}$, such that only one cone is visible.
		}
	}
	\label{fig:lateralDependence}	
\end{figure*}
%------- FIGURE -------

Several authors have studied cleaved surfaces of (Pb,Sn)Se with ARPES.\cite{Dziawa2012,Wojek2013,Pletikosic2014,Wojek2014,Neupane2015,Wojek2015} After accounting for differences in composition and temperature, the dispersion of the Dirac-like states is quite consistent. Scanning tunneling microscopy observations show that these cleaved surfaces consist of atomically flat terraces separated by 50-100~nm.\cite{Gyenis2013,Sessi2016} Since a typical ARPES light spot is of order $10 \mu$m $\times 10 \mu$m, it is interesting to consider why atomic steps do not influence the valley splitting of cleaved surfaces but only our PbSe overgrown surfaces. In Figure \ref{fig:lateralDependence}a we summarize tight binding calculations for a Pb$_{0.7}$Sn$_{0.3}$Se surface with 0.5~ML terrace coverage, where the lateral size of the terraces are varied (Fig.\ref{fig:lateralDependence}a). Note that this is a homostructure, not PbSe terraces on Pb$_{0.7}$Sn$_{0.3}$Se. We anticipate similar results in both cases.

Figure \ref{fig:lateralDependence}a reveals a size dependence of the splitting cancellation: for narrow terraces ($\leq$100 atomic rows, or 21~nm) the valley splitting has collapsed. For wider terraces the splitting is restored, together with a flat band connecting the two Dirac points. The 1D edge state was recently studied by Sessi \textit{et al.}.\cite{Sessi2016} For comparison, Fig.\ref{fig:lateralDependence}b shows flat surfaces with the same unit cell sizes as in Fig.\ref{fig:lateralDependence}a. This comparison demonstrates that the many bands appearing in Fig.\ref{fig:lateralDependence} are due to a bandfolding artifact from the large unit cells. After unfolding the bands only one Dirac cone would be observed in each panel.

In the effective mass model for valley splitting (see Supplementary Material) it is assumed that the wavefunction phase is fixed throughout the entire region of interest, in which case a surface half covered by $\pi$ phase shifting terraces should integrate to zero splitting. If the terrace separation is sufficiently large, the assumption of a fixed phase is no longer valid, \textit{i.e.} the wavefunction phase is able to adjust between terraces. This is the reason that the 1D edge state and the collapse of the Dirac cone splitting cannot occur simultaneously: the former results from the $\pi$ wavefunction phase difference at the boundaries of odd-height steps, while the latter occurs only when the wavefunction phases are unable to make a $\pi$ phase adjustment between surface steps.

Since cleaved surfaces consist primarily of large terrace widths of 50-100nm (230-465 atoms), the valley splitting is unaffected. For the PbSe overgrown surfaces studied here we expect a much higher terrace density. The small lattice mismatch between the (Pb,Sn)Se substrate and the PbSe layer should result in pseudomorphic growth. Meanwhile the very low growth temperature (close to room temperature) should suppress adatom mobility, causing layer-by-layer growth initiated at many disconnected islands. Previous work on IV-VI MBE on lead-salt substrates has demonstrated clear oscillations in reflection high energy electron diffraction (RHEED) down to the lowest studied substrate temperature of 70\textcelsius, a clear indication of a layer-by-layer growth mode.\cite{Springholz1994} 

We thus believe that a dense network of odd-height atomic steps is reponsible for causing the valley splitting collapse, and that these arise due to the low temperature, layer-by-layer partial filling of the large, atomically flat terraces on the cleaved substrate crystal. Such a growth mode would result in both a uniform height distribution and the necessary higher lateral density of steps compared with the cleaved substrate. Future studies combining topographic measurements (\textit{e.g.} scanning tunnelling microscopy or atomic force microscopy) with ARPES would be valuable to confirm this. 

Finally we comment on a potential implication of our observations. Previous studies \cite{Liu2014_nmat,Ezawa2014} have highlighted that 1D edge states in thin (001) oriented films of (Pb,Sn)Se also exhibit degeneracies shifted away from time reversal invariant momenta, and that this could be exploited for electronic devices. Since mirror symmetry can easily be broken by applying an electric field, an electrostatic gate could open a gap at the Fermi level. Our measurements highlight that while the interface degeneracies are robust, the valley interactions responsible for shifting them away from time reversal invariant momenta is fragile and depends sensitively on the atomic-scale condition of the surface. Presumably the same is true of the condition of edges in films. Without these valley interactions, all degeneracies occur as Kramers doublets. Although the remaining degeneracies would not be topologically protected if mirror symmetry were broken (owing to an even number of band inversions), still a substantial rearrangement of the bands would be required to eliminate all Fermi level crossings. It is not clear that the electric field produced by a gate electrode could accomplish this, in which case the proposed device would no longer act as a switch. Atomic-scale control of interfaces is fortunately within the capabilities of modern fabrication techniques, and may prove necessary to realize these device concepts.

%----------------------
\subsection*{Quantized PbSe States}
%----------------------
%------- FIGURE -------
\begin{figure}
	\includegraphics{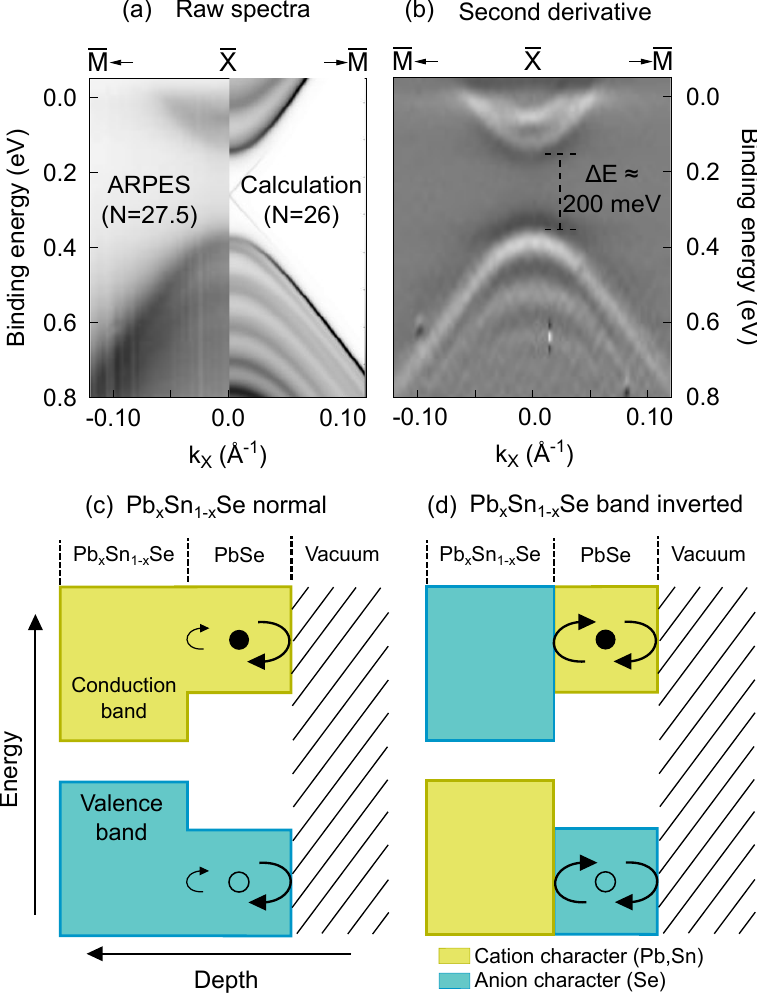}
	\caption{
		\textbf{Subband resonances in the thick film limit. With 27.5~ML of PbSe, the topological interface states are no longer observed. Instead the bandstructure (a) resembles that of bulk PbSe, including a bandgap $\Delta$E=200~meV . (b) A second derivative image corresponding to (a) highlights the existence of subband resonances in the conduction and valence band. When the substrate is band inverted, reflection of carriers at the PbSe / Pb$_{0.7}$Sn$_{0.3}$Se interface is enhanced due to the differing orbital character (c,d).  
		}
	}
	\label{fig:PbSeQW}	
\end{figure}
%------- FIGURE -------

Finally, we examine the outcome when thicker layers of PbSe are grown, shown in Fig.\ref{fig:PbSeQW}. With 27.5~ML (8.4~nm), the topological interface states are now buried beyond the probing depth of UV photoemission. As anticipated, the ARPES spectrum now resembles that of trivial PbSe at $\approx$130~K, including a 200~meV bandgap.\cite{Wojek2014} Careful inspection reveals subband resonances in both the valence and conduction bands, made clearer in a smoothed second derivative image (Fig.\ref{fig:PbSeQW}b). Spectral density calculations for a similar PbSe overlayer thickness (Fig.\ref{fig:PbSeQW}a, right hand side) reproduce these resonances.

Given that the Pb$_{0.7}$Sn$_{0.3}$Se substrate has a smaller bandgap ($\approx$50 meV \cite{Wojek2014}), a band alignment such as that sketched in Fig.\ref{fig:PbSeQW}c would not suggest strong confinement. However carriers in the PbSe layer still experience partial quantum mechanical reflection from the potential step at the PbSe / Pb$_{0.7}$Sn$_{0.3}$Se interface. The subband resonances would become more pronounced if the substrate were band inverted, as depicted in Fig.\ref{fig:PbSeQW}d and corresponding to the spectra in Fig.\ref{fig:PbSeQW}a,b. Now the anion/cation orbital character of the bands is reversed,\cite{Hsieh2012} and reflection of carriers at the interface is greatly enhanced (recall that the topological interface state exists because the two materials cannot otherwise electronically couple). This effect should be generic to any topological insulator / normal heterostructure with a sufficiently thin normal layer, and localized in momentum space to the vicinity of the band inversion. An immediate testable implication is that the subband resonances should be weakened by restoring a normal band ordering in the substrate, for example by applying heat or tensile strain. In this way, studying quantization in the overlayer may offer a remote probe of the buried topological interface.

%----------------------
\section*{CONCLUSIONS} 
%----------------------

In summary, by growing thin epitaxial films of the normal semiconductor PbSe on the TCI Pb$_{0.7}$Sn$_{0.3}$Se and studying these \textit{in-situ} with ARPES, we have gained a number of important insights. We have shown that the topological interface state is preserved when encapsulated in a heterostructure, and that ARPES observations of these buried interface states remain possible at depths of at least 8~ML. The valley splitting responsible for the \lq{}double Dirac cone\rq{} band dispersion is less robust, to the extent that the splitting can be eliminated by non-integer PbSe layer coverage. All of these findings have important implications for future efforts to build electronic devices from topological crystalline insulator materials. Our observations are also relevant as experimental manifestations of valley splitting, a phenomenon that is crucial to the coherent operation of spin-based solid state quantum computer schemes in multi-valley hosts such as Ge and Si.\cite{Goswami2007,Zwanenberg2013} Finally, we have observed quantized PbSe subbands for a layer thickness of 8.4~nm, and propose that this involves contributions from a type of symmetry confinement. Taken together, our experiments demonstrate that the approach of growing normal overlayers on topological substrates, to date largely unexplored, exposes a wide range of phenomena for further study.

\section*{Methods}

\textbf{Heterostructure Fabrication.} The substrates for these experiments were single crystal Pb$_{0.7}$Sn$_{0.3}$Se, grown \textit{ex-situ} by the self-selecting vapor-growth method \cite{Szczerbakow1994} and cleaved in ultrahigh vacuum to expose a clean (001) surface. All ARPES measurements were performed at a sample temperature of $\approx$130~K, bringing the substrates well into the TCI phase. For PbSe growth the samples were allowed to warm to room temperature, after which PbSe was deposited \textit{in-situ} by an effusion cell loaded with single crystal PbSe, similar to a previously described arrangement.\cite{Polley2014} Since PbSe evaporates molecularly,\cite{Springholz2007} stoichiometry is expected to be preserved in the grown layers. Substrates were not purposely heated during or after growth, but experienced radiative heating from the effusion cell. Low energy electron diffraction measurements confirmed that the PbSe overlayer retained the (001) orientation of the substrate, with no indication of a reconstruction. Despite the lack of substrate heating, the PbSe layers remained epitaxial and of relatively good quality, as demonstrated by the well defined bandstructures in Figures \ref{fig:Gap} and \ref{fig:PbSeQW}.

\textbf{Photoemission Measurements.}
Photoemission measurements were performed at the I4 beam line in the MAX-IV synchrotron facility,\cite{Jensen1997} using $18$~eV (ARPES) or $90$~eV (core level spectroscopy) p-polarized photons. The sample temperature was maintained at $\approx$130~K during measurements. For the ARPES spectra the total energy resolution was approximately 25~meV, with crystal momentum resolution better than $0.02$~\AA$^{-1}$. Additional measurements were performed at the SGM-3 beamline in the ASTRID 2 synchrotron facility; these are described in the Supplementary Material.

\textbf{Tight Binding Calculations.}
We employ a nearest neighbour 18-orbital \textit{sp$^3$d$^5$} model, with PbSe parameters taken from Lent \textit{et al.}.\cite{Lent1986} For Pb$_{0.7}$Sn$_{0.3}$Se we have applied the virtual crystal approximation and SnSe parameterization described previously.\cite{Safaei2015} We assume that the temperature dependence is dominated by the change of the lattice constant $a_0$, and rescale our parameterization according to the Harrison rules.\cite{Wojek2013} Similarly, biaxial strain is treated by rescaling hopping parameters such that the change in energy gap with strain is in agreement with the experimentally determined deformation potential \cite{Simma2009} and elastic constants.\cite{Lippmann1971} The spectral densities projected on specific atomic layers have been calculated with the use of a tight-binding Hamiltonian for a semi-infinite system and an appropriate modification of a recursive Green's function method described by Lopez Sancho.\cite{LopezSancho1985}

\section*{Supporting Information Available}
Details of Dirac point valley splitting in the effective mass model, additional photoemission measurements, fitting of the ARPES spectra to a $k \cdot p$ model, details of core level photoemission analysis, \textit{ex-situ} cross sectional electron microscopy and secondary ion mass spectrometry. 

\section*{Author Contributions}
C.M.P., A.F., M.B, A.G.{\v C} and T.B. performed the PbSe deposition and photoemission measurements. A.S. grew the single crystal (Pb,Sn)Se substrates. P.D. and M.T. performed \textit{ex-situ} analysis of the samples. R.B. developed the effective mass and tight binding models, and with R.R performed calculations. C.M.P. carried out the data analysis and wrote the manuscript, with input from all authors. All authors contributed to the interpretation.

\section*{Acknowledgements}
C.M.P. thanks T.B. Boykin and M.A. Eriksson for helpful discussions, and J. Osiecki for creating data analysis software used here. This work was made possible through support from the Knut and Alice Wallenberg Foundation, the Swedish Research Council, the VILLUM FONDEN \textit{via} the Centre of Excellence for Dirac Materials (Grant No. 11744) and the Polish National Science Centre (NCN) under projects 2013/11/B/ST3/03934, 2014/15/B/ST3/03833 and 2014/15/B/ST3/04489.

\end{document}

\author{Craig M. Polley}
\email{craig.polley@gmail.com}
\affiliation{MAX IV Laboratory, Lund University, 221 00 Lund, Sweden}
\author{Ryszard Buczko}
\affiliation{Institute of Physics, Polish Academy of Sciences, 02-668 Warsaw, Poland}
\author{Alexander Forsman}
\affiliation{KTH Royal Institute of Technology, SCI Materials Physics, S-164 40 Kista, Sweden}
\author{Piotr Dziawa}
\author{Andrzej Szczerbakow}
\author{Rafa\l{} Rechci\'nski}
\author{Bogdan J. Kowalski}
\author{Tomasz Story}
\affiliation{Institute of Physics, Polish Academy of Sciences, 02-668 Warsaw, Poland}
\author{Ma\l{}gorzata Trzyna}
\affiliation{Center for Microelectronics and Nanotechnology, Rzeszow University, Rejtana 16A, Rzeszow 35-959, Poland}
\author{Marco Bianchi}
\author{Antonija Grubi{\v s}i{\'c} {\v C}abo}
\author{Philip Hofmann}
\affiliation{Department of Physics and Astronomy, Interdisciplinary Nanoscience Center (iNANO), Aarhus University, 8000 Aarhus C, Denmark}
\author{Oscar Tjernberg}
\affiliation{KTH Royal Institute of Technology, SCI Materials Physics, S-164 40 Kista, Sweden}
\author{Thiagarajan Balasubramanian}
\affiliation{MAX IV Laboratory, Lund University, 221 00 Lund, Sweden}

\title{Supplementary Information for `Fragility of the Dirac Cone Splitting in Topological Crystalline Insulator Heterostructures'}

\maketitle
\onecolumngrid

%------------------------------------------------------------------
\section{Dirac Point Valley Splitting in the Presence of Surface Terraces}
%------------------------------------------------------------------
The $( 0 0 1 )$ surface of Pb$_{1-x}$Sn$_{x}$Se breaks the symmetry of translation along the $[0 0 1]$ direction, permitting the existence of surface states composed of the contributions from two $L$ valleys at the $L_1=(1,1,1)\frac{\pi}{a_0}$ and $L_2=(1,1,-1)\frac{\pi}{a_0}$ points in the bulk Brillouin zone.

Let us assume that we know the solutions of the effective mass equation for the nontrivial surface or interface states obtained only for one valley $L_1$. We restrict our consideration only to the solution at $\overline{X}$ in the surface Brillouin zone, and to eigenstates of the $(1\overline{1}0)$ mirror plane which correspond to eigenvalue $M=-i$. After omitting quadratic $k$ terms in the Hamiltonian, one can obtain a wavefunction of the form:
\begin{equation}
\label{F1}
F_1(\overrightarrow{r})= e^{i\varphi}f(z)(a\Psi_1^+(\overrightarrow{r}) + b\Psi_1^-(\overrightarrow{r}))
\end{equation}
where $a$ and $b$ are real and $\Psi_1^\pm$ are the Bloch functions transforming according to $L_6^+$ and $L_6^-$ irreducible representations of the $D_{3d}$ point group, as known for IV-VI semiconductors. The envelope real function $f(z)$ is exponentially decaying for $z \rightarrow \infty$ (substrate bulk region). The phase $\varphi$, in the general case of an anisotropic effective mass, depends on $z$ and is defined up to an arbitrary phase $\varphi_0$. 

The appropriate wavefunction for the $L_2$ valley can be obtained by time reversal symmetry $\Theta$ followed by the $\pi$ rotation $C_2$ around the $[110]$ axis.
\begin{equation}
\label{F2}
F_2(\overrightarrow{r})=C_2 \Theta F_1(\overrightarrow{r})= e^{-i\varphi}f(z)(a\Psi_2^+(\overrightarrow{r}) + b\Psi_2^-(\overrightarrow{r}))
\end{equation}

The valley mixing effect can be described by a simple model proposed by Liu \cite{Liu2013_prb}, where the zero order mixing operator in the $M=-i$ subspace is given by:
\begin{equation}
\label{Delta}
\Delta=m\tau_x + \delta \tau_y
\end{equation}
Within this model the antibonding and bonding superpositions of $F_i$ are:
\begin{equation}
F_{\pm}=(F_1 \pm e^{i\chi}F_2)/\sqrt{2}
\end{equation}
where $\chi=\arg( m+i\delta )$. The phase $\varphi_0$ is no more arbitrary, because it alters the phase difference between two components in $F_{\pm}$, and we have assumed that $\varphi_0=0$ for above special solutions corresponding to two Dirac points. The valley splitting is given by $2\sqrt{m^2+\delta^2}$.

The $\Psi_{1,2}^{\pm}$ Bloch wavefunctions in equations \ref{F1},\ref{F2} are combinations of the Bloch sums defined by atomic-like (L$\ddot{o}$wdin) orbitals:
\begin{equation}
\Psi_i^\pm=\sum_{\alpha}c_{\alpha,i}^\pm \Phi_{\alpha,i}
\end{equation}
where $\alpha$ denotes cation or anion orbitals associated with the appropriate spin aligned in the direction perpendicular to the mirror plane $(1,-1,0)$, and

\begin{equation}
\Phi_{\alpha,i}(\overrightarrow{r})=\frac{1}{\sqrt{N}} \sum_{\overrightarrow{t}} e^{i\overrightarrow{L}_i (\overrightarrow{t}+\overrightarrow{d})}\\
         \phi_{\alpha}(\overrightarrow{r}-\overrightarrow{t}-\overrightarrow{d})
\end{equation}
The translation vectors
$\overrightarrow{t}=l\overrightarrow{t_1}+m\overrightarrow{t_2}+n\overrightarrow{t_3}$
are based on the primitive lattice translation vectors:

$\overrightarrow{t_1}=(1,-1,0)\frac{a_0}{2}$,
$\overrightarrow{t_2}=(1,1,0)\frac{a_0}{2}$ and
$\overrightarrow{t_3}=(1,0,1)\frac{a_0}{2}$.

Note that $\overrightarrow{t_1}$ and $\overrightarrow{t_2}$ are parallel to (001) surface.
\\
\\
The displacement vectors $\overrightarrow{d}$ are fixed offsets according to the lattice structure: 
$\overrightarrow{d}=0$ for cation and $\overrightarrow{d}=(1 0 0)\frac{a_0}{2}$ for anion atom positions.
\\
\\
\\
Rather than using the $\Phi_{\alpha,i}$ basis, it is more convenient to work with combinations:

\begin{equation}
\Phi_{\alpha,+}=(\Phi_{\alpha,1}+\Phi_{\alpha,2})/\sqrt{2},\qquad \Phi_{\alpha,-}=(\Phi_{\alpha,1}-\Phi_{\alpha,2})/\sqrt{2}
\end{equation}

The appropriate Bloch sums are now:
\begin{eqnarray}
\Phi_{\alpha,+}(\overrightarrow{r})=\frac{1}{\sqrt{N}} \sum_{l,m} (-1)^m \sum_{j}
         \phi_{\alpha}(\overrightarrow{r}-l\overrightarrow{t_1}-m\overrightarrow{t_2}-2j\overrightarrow{t_3}-\overrightarrow{d})
         \nonumber \\
\Phi_{\alpha,-}(\overrightarrow{r})=\frac{-1}{\sqrt{N}} \sum_{l,m} (-1)^m \sum_{j}
         \phi_{\alpha}(\overrightarrow{r}-l\overrightarrow{t_1}-m\overrightarrow{t_2}-(2j+1)\overrightarrow{t_3}-\overrightarrow{d})
\end{eqnarray}
\\
\\
An important aspect of the above sums is that they are combinations of orbitals centered at every second atomic plane, as a direct consequence of the fact that the $L$ point separation is equal to $2\pi/a_0[0,0,1]$. If we define the index $j$ as equal to one at the (001) surface plane, then $\Phi_{-}$ is defined by atoms at the surface and all odd numbered atomic planes, whereas $\Phi_{+}$ contains atoms situated in even numbered planes.
\\
\\
After covering the surface with a new atomic layer, the role of $\Phi_{-}$ and $\Phi_{+}$ is exchanged. Now, with $j$ starting from 0 in the sums, $\Phi_{+}$ is defined by surface atoms and $\Phi_{-}$ by atoms in the next layer. It can be shown that also the antibonding and bonding character of $F_\pm$ combinations is exchanged, and valley splitting remains unaffected. It should be noticed that the exchange of $F_\pm$ character can be also achieved by changing $\varphi_0$ to $\pi/2$, which leads a change of the total phase difference between two $L$ valley components by $\pi$.
\\
\\
In the case of a surface covered by a sufficiently dense array of terraces or steps, the $\varphi_0$ remains fixed across the entire surface. But since odd-height regions cancel the valley interactions produced by surface and even-height regions, the overall effect will be a cancellation of the valley splitting, dependent on the ratio of odd- to even-height regions on the surface. In the case of terraces made from the same atoms as the substrate (e.g. homoepitaxial growth), total cancellation should occur for half coverage of odd-height terraces. When terraces are made with different material, the assumption typically used in effective mass theory of identical Bloch functions across the entire structure is not well satisfied, and the coverage for total cancellation can differ from 1/2.
\\
\\
Each of the $F_{\pm}$ solutions is constructed with Bloch sums corresponding to even and odd atomic layers, however for a given orbital $\alpha$ the contributions of $\Phi_{\alpha,-}$ and $\Phi_{\alpha,+}$ are not the same. They depend on $\alpha$ and are opposite for bonding and antibonding states. As a result $F_\pm$ wavefunctions rapidly oscillate in antiphase and with two atomic layer periodicity, as shown in Fig. 5d. This leads to another form of valley splitting oscillation with overlayer thickness when the wavefunctions are localized at the interface between a TCI and normal insulator and the surface layer is always fully completed.
\\
\\
\newpage
%------------------------------------------------------------------
\section{Supporting Experiments}
\begin{figure*}[h]
	\includegraphics[width=16cm]{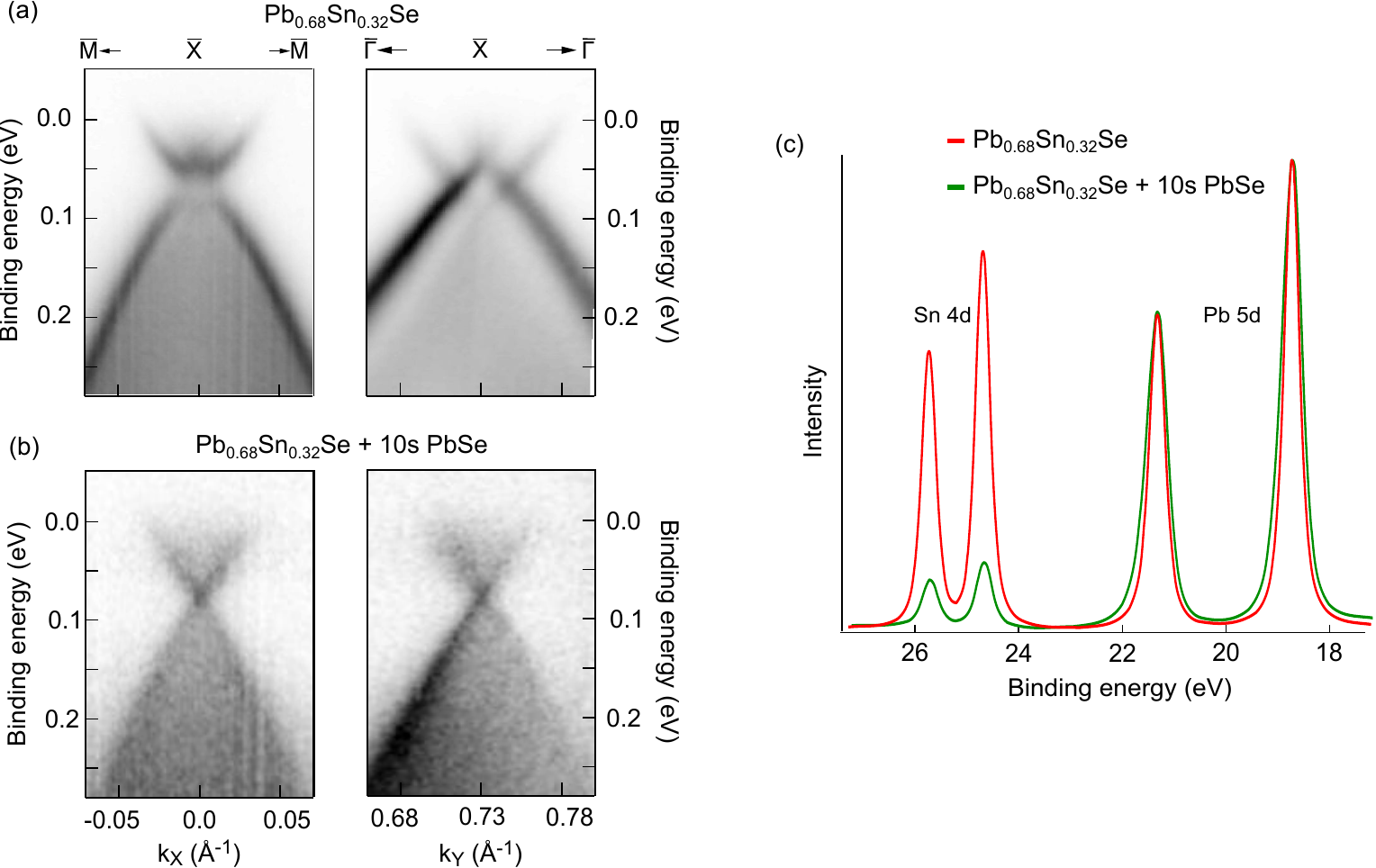}
	\caption{
		\textbf{Measurements taken at the i4 beamline at the MAX-IV facility on a substrate with slightly higher Sn content (x=0.32), confirming the collapse of the valley splitting after PbSe deposition. The sample temperature was maintained at $\approx$130~K during measurements using a bath liquid nitrogen cryostat.
		}
	}
	\label{fig:A1}	
\end{figure*}

\begin{figure*}[h]
	\includegraphics[width=16cm]{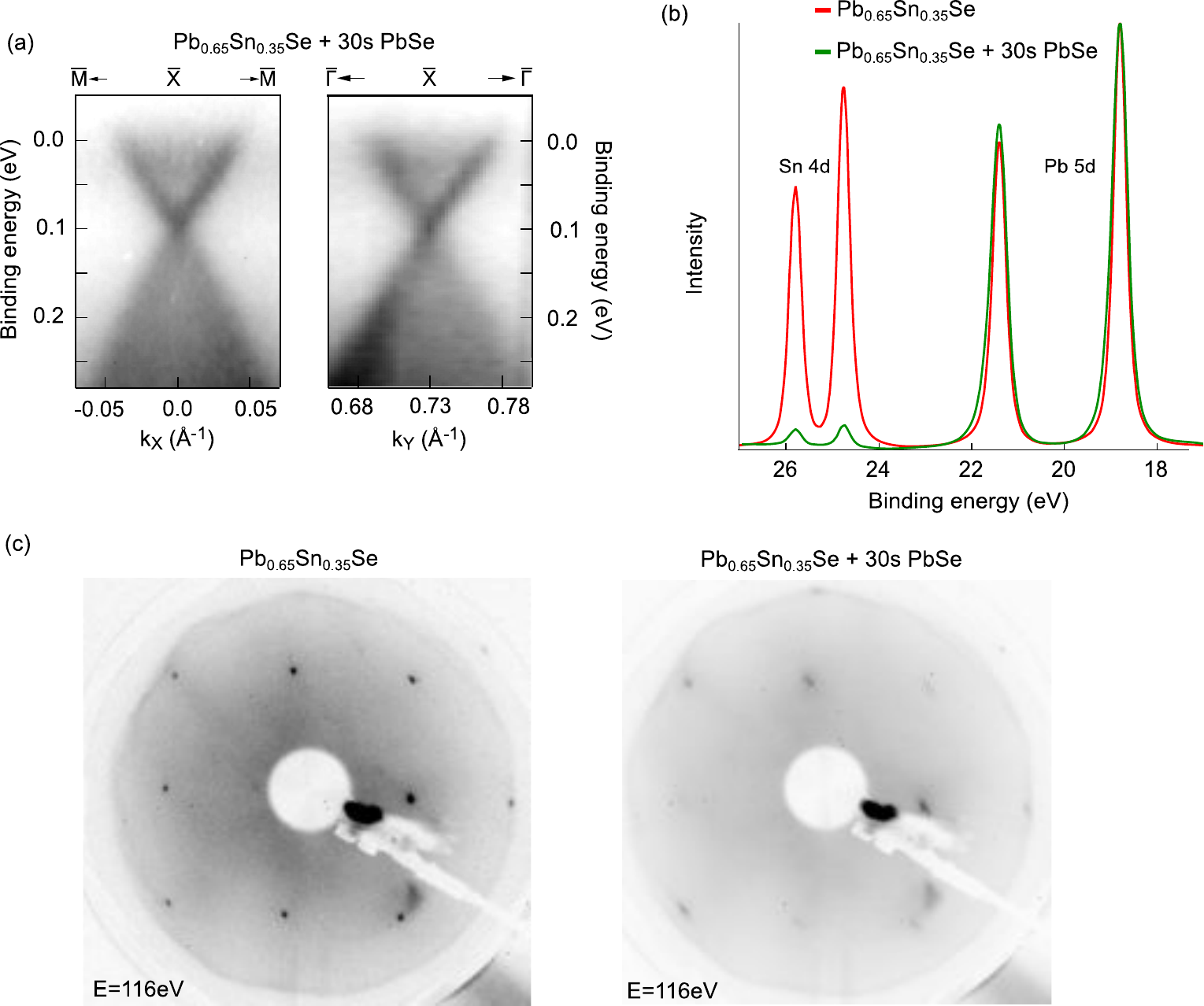}
	\caption{
		\textbf{Measurements taken at the i4 beamline at the MAX-IV facility on a substrate with slightly higher Sn content (x=0.35), confirming the collapse of the valley splitting. The sample temperature was maintained at $\approx$130~K during measurements using a bath liquid nitrogen cryostat. LEED measurements before and after PbSe deposition (c) demonstrate that the PbSe overlayer retains the (001) orientation of the substrate without reconstruction. 
		}
	}
	\label{fig:A2}	
\end{figure*}

\begin{figure*}[h!]
	\includegraphics[width=16cm]{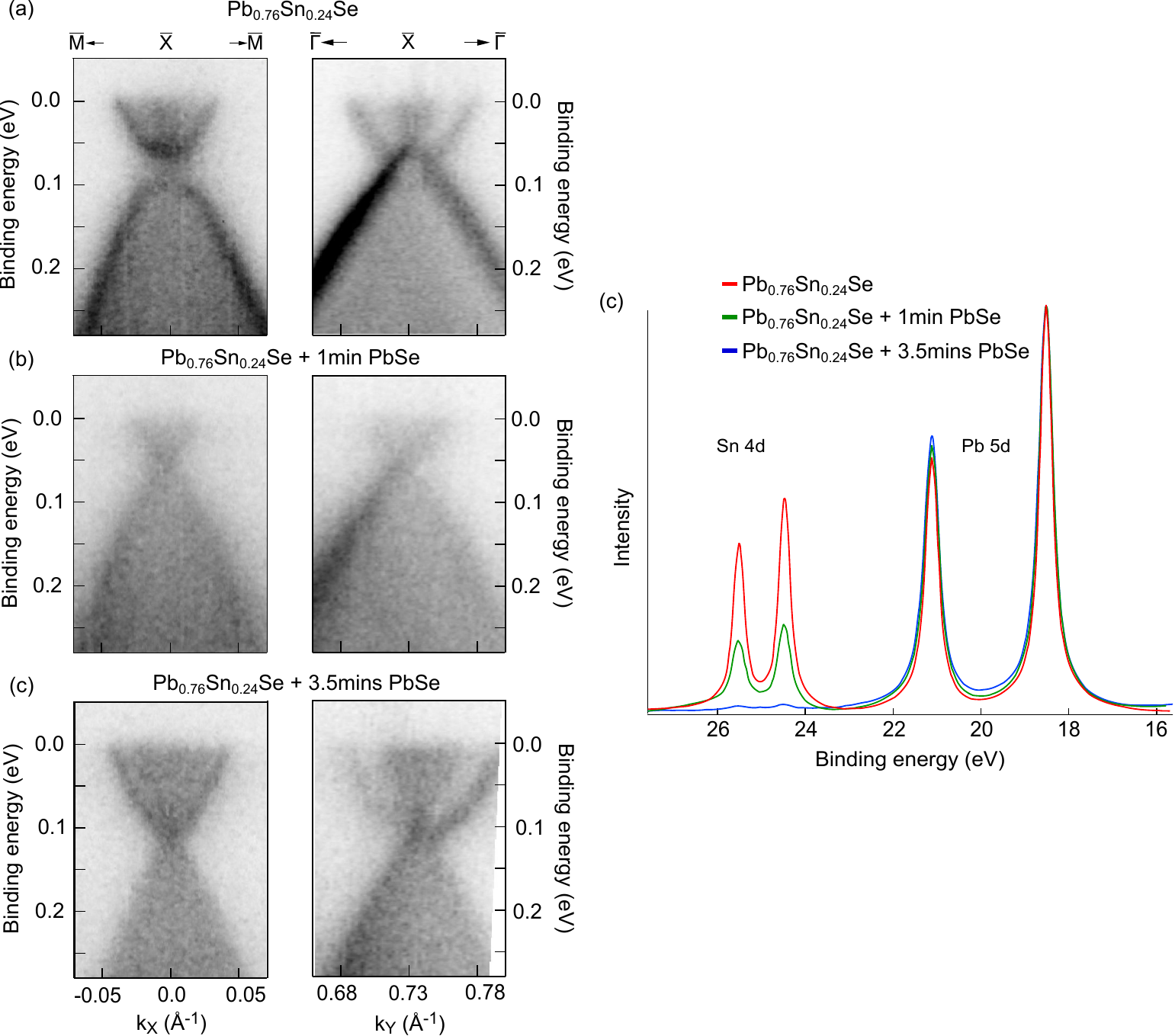}
	\caption{
		\textbf{Measurements taken at the SGM-3 beamline at the ASTRID 2 facility on a substrate with lower Sn content (x=0.24), confirming the oscillatory changes with thickness. PbSe growth was performed with the same PbSe source but with higher substrate temperatures ($\approx$70\textcelsius). The sample temperature was maintained at $\approx$16~K during measurements using a liquid helium cryostat.
		}
	}
	\label{fig:A3}	
\end{figure*}

%------------------------------------------------------------------
In Figs.\ref{fig:A1}-\ref{fig:A3} we summarize additional sample preparations and measurements. While the growth conditions were broadly similar to those employed in the main manuscript, these measurements lack a reliable thickness characterization. While a comparable growth rate is expected (2 ~ML min$^{-1}$), we specify only the PbSe deposition time. These measurements demonstrate that the changes in Dirac cone splitting are reproducible. All spectra were acquired with p-polarized photons, with photon energies of $18$~eV (ARPES) or $90$~eV (core levels).

%------- FIGURE -------
%------- FIGURE -------
\newpage
%------- FIGURE -------
%------- FIGURE -------
\newpage
%------- FIGURE -------
%------- FIGURE -------
\newpage
%----------------------
\section{$k \cdot p$ Fitting and Changes in Dispersion}
%----------------------
On the clean surface the two parent Dirac cones have identical k$_\parallel$ dispersion, but careful inspection reveals that this is not the case after 3.9~ML of PbSe deposition (Fig. 4g). Such behaviour is included in the symmetry derived $k \cdot p$ model of Liu \cite{Liu2013_prb}, provided we include higher order terms:
\begin{equation}\label{eqn:F1}
H(k) = (v_x k_x s_y - v_y k_y s_x) + (v'_x k_x s_y - v'_y k_y s_x) \tau_x +  m \tau_x + \delta s_x \tau_y
\end{equation}
Here $x$ is the $\bar{X}$-$\bar{M}$ direction and $y$ the $\bar{X}$-$\bar{\Gamma}$ direction. The parameter $v$ is the band velocity, modified for each parent cone by $v'$ (one becomes $v+v'$, the other $v-v'$), $m$ is the energy splitting of the parent Dirac cones at $k=\bar{X}$ and $\delta$ is the hybridization strength between the parent Dirac cones outside of the mirror symmetry protected $\bar{\Gamma}$-$\bar{X}$-$\bar{\Gamma}$ plane. The matrices $s$ and $\tau$ are Pauli matrices in spin and valley space. All but the first two terms in Eqn. \ref{eqn:F1} are off-diagonal in valley space (through $\tau_x$ and $\tau_y$), and thus introduce effects of intervalley interactions.

This expanded Hamiltonian can be solved numerically for the band dispersions, but rigorously fitting our experimental data with this model is not straightforward. The real spectra contain diffuse intensity from other sources such as scattering from nearby bulk bands, which the 4-band $k \cdot p$ model does not describe. In Table \ref{table:kp_parameters} we summarize our best estimates of the $k \cdot p$ parameters for the three clearest datasets. Peak positions extracted from momentum and electron distribution curves yield the band velocity parameters $v$ and $v'$, and by extrapolation a first estimate of the energy gap $\Delta$. A non-zero hybridization term $\delta$ results in bending of the bands which slightly enlarges the apparent energy splitting $m$. This coupling implies that it is insufficient to analyze only a $\bar{\Gamma}$-$\bar{X}$-$\bar{\Gamma}$ image. In addition, when $m$=0 no band crossings occur outside of the mirror plane, so $\delta$ cannot be extracted. Here we determine $m$ and $\delta$ by iteratively calculating the band dispersion and comparing the output to the $\bar{M}$-$\bar{X}$-$\bar{M}$ and $\bar{\Gamma}$-$\bar{X}$-$\bar{\Gamma}$ spectra.

%------- FIGURE -------
\begin{figure}[h]
	\includegraphics[width=8cm]{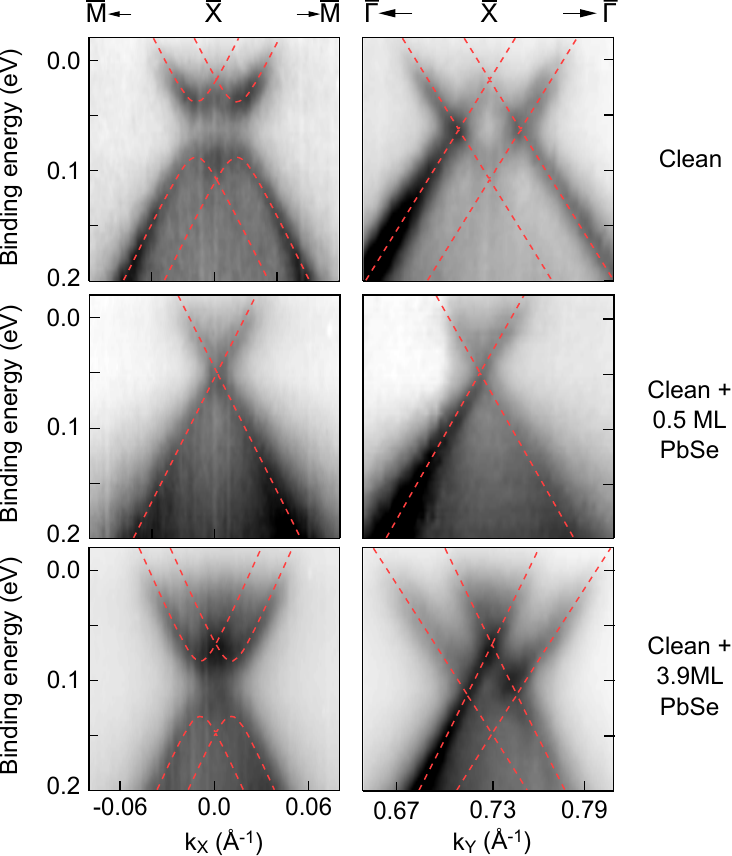}
	\caption{
		\textbf{Theoretical description of the k$_{//}$ dispersions. Applying the symmetry derived $k \cdot p$ model model of Liu \cite{Liu2013_prb} (red dashed lines) using empirically determined parameters listed in Table \ref{table:kp_parameters} captures the evolving behaviour of the interface state bandstructure.
		}
	}
	\label{fig:kp}	
\end{figure}
%------- FIGURE ------- 
\begin{table}[h]
\centering
\begin{tabular}{|c|c|c|c|c|c|c|}
\hline
\parbox[t]{1.65cm}{PbSe depth\\(ML)} & \parbox[t]{1cm}{$v_x$\\(eV\AA)} & \parbox[t]{1cm}{$v_y$\\(eV\AA)} & \parbox[t]{1cm}{$v'_x$\\(eV\AA)} 	& \parbox[t]{1cm}{$v'_y$\\(eV\AA)} & \parbox[t]{1cm}{$m$\\(meV)} 	& \parbox[t]{1cm}{$\delta$\\(meV)}\\
\hline
0 					& 2.9 			& 2.3 			& 0			&	0		& 76				& 25\\ %Final!
\hline
0.5 				& 2.8 			& 2.5 			& 0			&	0		& 0					& -- \\	%Final!
\hline
3.9					& 3.2 			& 2.6 			& 0			&	0.4		& 64				& 25 \\	%Final!
\hline
\end{tabular}
\caption{Parameters used in the $k \cdot p$ model to generate the curves in Figure \ref{fig:kp}.}
\label{table:kp_parameters}	
\end{table} 
With very little PbSe coverage, the inter-valley interactions appear to vanish. The resulting band dispersion is nearly isotropic, which can be understood from consideration of the bulk bandstructure. In contrast to (Pb,Sn)Te, the bulk conduction and valence bands in (Pb,Sn)Se have nearly isotropic effective masses. With respect to the major axis along the [111] direction, the effective mass anisotropy $\frac{m_\parallel}{m_\perp}$ begins at 1.7 for PbSe and approaches unity for Sn contents $>$0.2 \cite{NimtzSchlicht1983}. Since these ellipsoids are tilted with respect to the (001) surface, the surface state effective mass anisotropy is further reduced to $\frac{1}{3} (2 + m_\parallel/m_\perp)$ \cite{Stern1967}.

Changes in the valley splitting term $m$ have already been discussed, but the non-zero $v'_y$ for the 3.9~ML sample warrants further comment. A band velocity adjustment could equivalently be interpreted as a $k_\parallel$-dependent valley splitting term. Since there have been no momentum-resolved experimental probes of valley splitting phenomena until very recently \cite{Miwa2014}, momentum dependence has had limited theoretical consideration \cite{Ting1988,Ikonic1990}. Since the interface state wavefunction is composed from the set of bulk Bloch states with the same $k_\parallel$, it is straightforward to see that changing $k_\parallel$ could modify this set and hence the valley splitting of the interface states.
\newpage
%------------------------------------------------------------------
\section{Core Level Analysis Details}
%------------------------------------------------------------------

At each PbSe depth, the Pb 5d and Sn 4d core level spectra were acquired with identical spectrometer settings. Peak fitting was performed with a single Voigt doublet for each species separately, which allowed the Shirley background factor to differ. For the analysis performed in this study, the area of the background and of the peaks are the most important aspects; issues such as multiple components or minor binding energy offsets are of little consequence.
\\
The peak model was calculated iteratively as:
\begin{enumerate}
\item Generate a Gaussian convolved Lorentzian doublet
\item Compute an active Shirley background from this doublet \cite{Gomez2014}
\item Add a constant offset
\item Iterate until $\chi ^2$ is minimized
\end{enumerate}

\textbf{Estimating Sn fraction:} Peak areas $A$ were determined by numerically integrating the doublet from the fitting model (background subtracted). The tin fraction can then be estimated by:
\begin{equation}
x = \frac{(\frac{A}{\sigma})_{Sn 4d}} {(\frac{A}{\sigma})_{Sn 4d} + (\frac{A}{\sigma})_{Pb 5d}}
\end{equation}

Where the cross sections at h$\nu$=90~eV are $\sigma_{Sn 4d}$=4.50~Mb and $\sigma_{Pb 5d}$=5.17~Mb, obtained from Elettra's online ``WebCrossSection'' service which is based on the calculations of Yeh and Lindau \cite{Yeh1985}

\textbf{Estimating PbSe thickness:} On the clean Pb$_{0.7}$Sn$_{0.3}$Se surface the photoemission intensity from the Sn 4d core level can be described as a summation over every slice in the vertical direction, attenuated according to the Beer-Lambert absorption law:
\begin{equation}
\begin{split}
I_{Sn 4d} 	&= K \sigma _{Sn 4d} N_{Sn} \int_0^{\infty} \exp\Big(-\frac{z}{\lambda\cos{\theta}}\Big) dz \\
			&= K \sigma _{Sn 4d} N_{Sn}\lambda\cos{\theta}
\end{split}
\end{equation}

Where $K$ is a scaling factor based on experimental conditions such as photon flux, sample alignment and analyzer settings, $\sigma$ is the photoionization cross section, $N_{Sn}$ the planar atomic density of Sn atoms in Pb$_{0.7}$Sn$_{0.3}$Se, $z$ the distance into the substrate from the surface and $\lambda$ the inelastic mean free path of the photoelectron. Since the measurements are made at an angle $\theta$=23° away from normal emission (corresponding to $\bar{X}$ at h$\nu$=18~eV), the attenuation length is extended by the factor $\frac{1}{\cos{\theta}}$.
\\
\\
When a PbSe layer of thickness $d$ is deposited on the Pb$_{0.7}$Sn$_{0.3}$Se, this same total intensity from the Sn is now attenuated by having to travel through the PbSe layer:
\begin{equation}
I_{Sn 4d}(d) = K \sigma _{Sn 4d} N_{Sn}\lambda \exp\Big(-\frac{d}{\lambda cos(\theta)}\Big)
\end{equation}

In practice the scaling factor $K$ differs between measurements, so that normalization is required. Here we normalize to the Pb 5d peak. On the clean Pb$_{0.7}$Sn$_{0.3}$Se surface this has an intensity given by: 
\begin{equation}
I_{Pb 5d} = K \sigma _{Pb 5d} N_{Pb}\lambda\cos{\theta}
\end{equation}

The PbSe overlayer has a higher atomic density of Pb compared to Pb$_{0.7}$Sn$_{0.3}$Se, equivalent to the sum $(N_{Pb}+N_{Sn})$ in Pb$_{0.7}$Sn$_{0.3}$Se (i.e. $1=0.7+0.3$). A simple way to incorporate this is to separate the overlayer contribution into that from $N_{Pb}$ and $N_{Sn}$ lead atoms, in which case we can write:
\begin{equation}
\begin{split}
I_{Pb 5d}(d) &= K \sigma _{Pb 5d}N_{Pb}\lambda\cos{\theta} + K*\sigma _{Pb 5d}N_{Sn}  \int_0^d \exp\Big(-\frac{z}{\lambda\cos{\theta}}\Big) dz \\
			 &=	K \sigma _{Pb 5d}N_{Pb}\lambda\cos{\theta} + K \sigma _{Pb 5d}N_{Sn}\lambda\cos{\theta}\Bigg(1-\exp\Big(-\frac{d}{\lambda\cos{\theta}}\Big)\Bigg)
\end{split}
\end{equation}

The depth dependent ratio of core level intensities $R(d)$ is given by:
\begin{equation}
\begin{split}
R(d) 	&= \frac{I_{Pb 5d}(d)}{I_{Sn 4d}(d)} \\
		&= \frac{\sigma _{Pb 5d}N_{Pb}}{\sigma _{Sn 4d}N_{Sn}\exp\big(-\frac{d}{\lambda cos(\theta)}\big)}  + \frac{\sigma _{Pb 5d}\Big(1-\exp\big(-\frac{d}{\lambda cos(\theta)}\big)\Big)}{\sigma _{Sn 4d}\exp\big(-\frac{d}{\lambda cos(\theta)}\big)} \\
		&=	\frac{R(0)}{\exp\big(-\frac{d}{\lambda cos(\theta)}\big)} + \frac{\sigma _{Pb 5d}}{\sigma _{Sn 4d}}\Big(\exp\big(\frac{d}{\lambda cos(\theta)}\big)-1\Big)
\end{split} \label{fit}
\end{equation}

\begin{figure*}[h]
	\includegraphics[width=8cm]{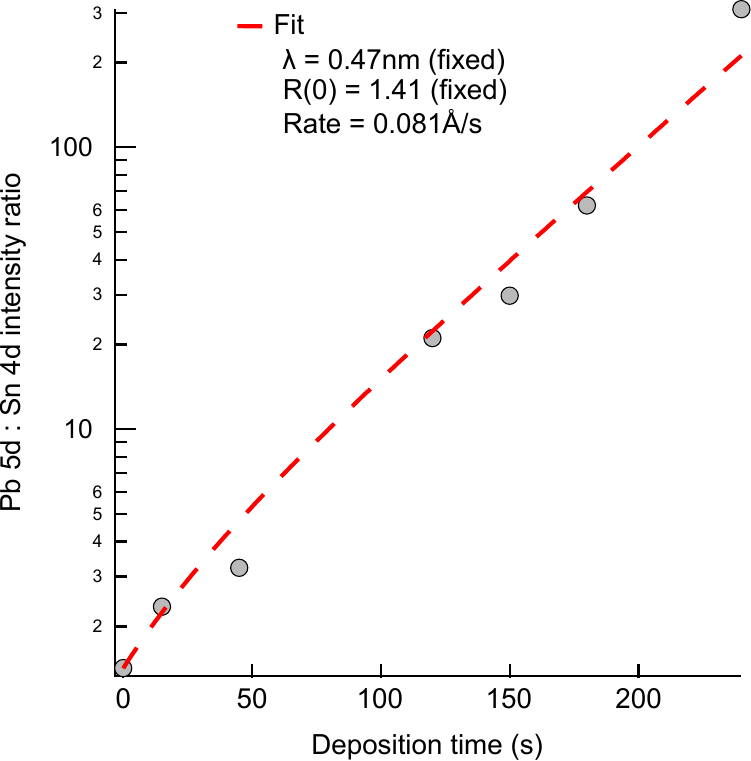}
	\caption{
		\textbf{Fitting measured core level intensity ratios with Equation \ref{fit} to extract a PbSe growth rate estimate.}
	}
	\label{fig:B1}	
\end{figure*}

Since $R(d)$ is experimentally measured, we can determine the growth rate by fitting the above expression to our measurements (substituting (growth rate$\times$time) for $d$). The inelastic mean free path in PbSe is not documented, but we expect it to be close to that of PbTe ($\lambda$=0.47~nm at E$_K$=63~eV, based on the ``TPP-2M'' formula of Tanuma, Powell and Pen \cite{Tanuma1991,Powell2010}). As shown in Fig.\ref{fig:B1}, this yields a growth rate of 0.08~\AA s$^{-1}$. The rate determined by \textit{ex-situ} electron microscopy should be considered more reliable, since the calculation here is strongly dependent on the parameter $\lambda$ and the accuracy of the TPP-2M result has not been quantified.

%------- FIGURE -------
%------- FIGURE -------
\newpage

%------------------------------------------------------------------
\section{\textit{Ex-Situ} Thickness Calibration}
\begin{figure*}[h]
	\includegraphics[width=8cm]{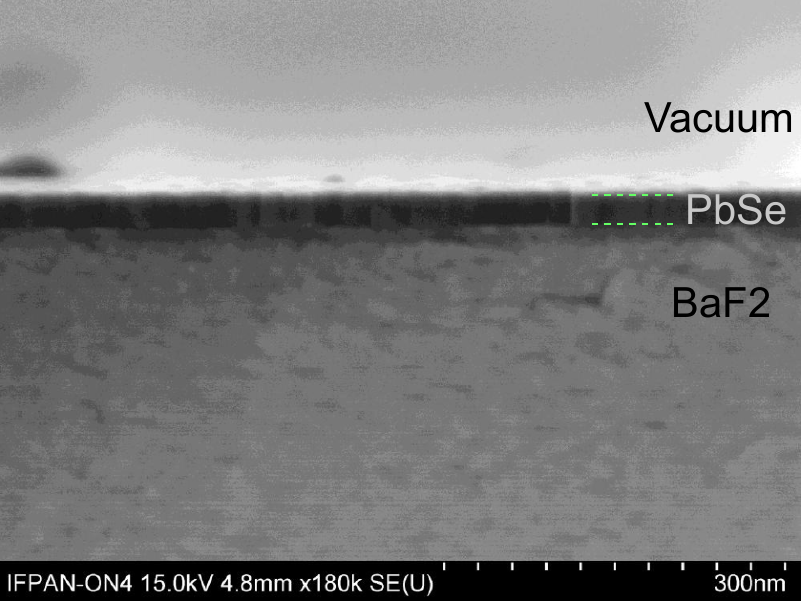}
	\caption{
		\textbf{Example of a cross sectional electron microscopy image of the thick calibration film of PbSe on BaF$_2$.}
	}
	\label{fig:C1}	
\end{figure*}

%------------------------------------------------------------------
\textbf{Cross Sectional Electron Microscopy.}
Immediately following the experiments described in the manuscript, PbSe was deposited on a clean  BaF$_2$ substrate for 46 minutes. This gives a thick film which can easily be measured by \textit{ex-situ} cross sectional electron microscopy (Fig.\ref{fig:C1}). Examining 4 different images we obtain a film thickness of (28.7 $\pm$ 3.0)~nm, equivalent to a growth rate of (0.10 $\pm$ 0.01)~\AA s$^{-1}$.

%------- FIGURE -------
%------- FIGURE -------

\textbf{Secondary Ion Mass Spectrometry.}
Time-of-flight secondary ion mass spectrometry (TOF-SIMS) measurements were performed on the sample from which the ARPES in the manuscript was measured, using a TOF SIMS 5 (IONTOF GmbH, M\"{u}nster, Germany) in dual beam mode. A Bi$^+$ primary beam (30 keV / 1.37 pA) and Cs$^+$ sputter beam (500 eV / 18 nA) were used in a non-interlaced mode. Positive secondary ions of PbCs$^+$, SeCs$^+$ and SnCs$^+$ were collected. The primary beam was rastered over an area of 100$\times$100 $\mu$m$^2$, centered inside a sputter crater of 300$\times$300 $\mu$m$^2$. The positive secondary ion mass spectra were calibrated using C, CH$_3^+$,Sn$^+$, Se$^+$, Cs$^+$ and Cs$_2^+$ Depth was calibrated based on crater depth, as measured by a DEKTAK stylus profilometer.

The interface positions were determined from the half maximum of the Pb and Se (front surface) and Sn (back surface) signals. This approach yields a film thickness of 9.4 nm, equivalent to a growth rate of 0.11~\AA s$^{-1}$. The Sn signal drops sharply at the growth interface without obvious signs of segregation, in accordance with the photoemission analysis in the manuscript.

%------- FIGURE -------
\begin{figure*}[h]
	\includegraphics[width=7.5cm]{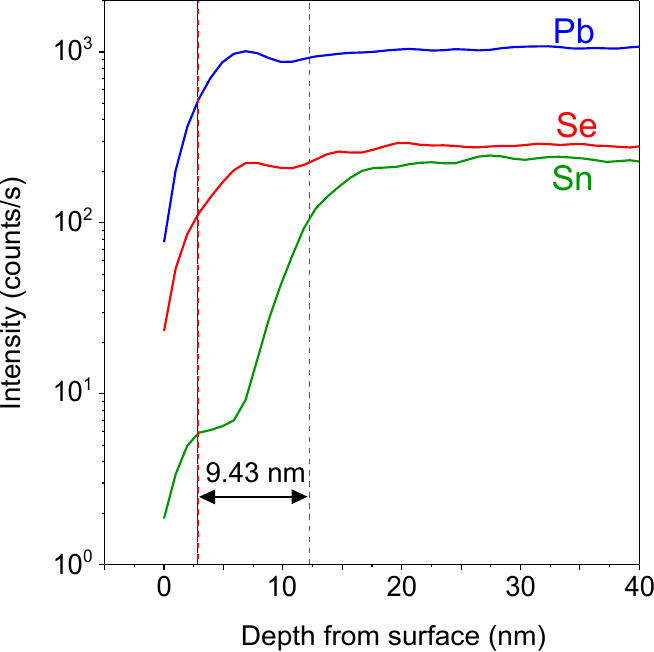}
	\caption{
		\textbf{TOF-SIMS depth profiles of the Pb,Se and Sn species in the sample discussed in the manuscript.}
	}
	\label{fig:C2}	
\end{figure*}
%------- FIGURE -------